\input harvmac
%\draftmode
\noblackbox
%-------------------------
% This paper uses harvmac
%-------------------------
\font\ticp=cmcsc10
 
\def\Title#1#2{\rightline{#1}\ifx\answ\bigans\nopagenumbers\pageno0\vskip1in
\else\pageno1\vskip.8in\fi \centerline{\titlefont #2}\vskip .5in}

\font\ticp=cmcsc10
\font\ttsmall=cmtt10 at 8pt

%
% definitions
%
\def\[{\left [}
\def\]{\right ]}
\def\({\left (}
\def\){\right )}

% references

\lref\more{See e.g. G. Gibbons and K. Maeda, Nucl. Phys. {\bf B298} (1988) 741;
P. Bizon, Phys. Rev. Lett. {\bf 64} (1990) 2844; K. Y. Lee, V. P. Nair and
E. Weinberg, Phys. Rev. Lett. {\bf 68} (1992) 1100.}
\lref\pois{E. Poisson and W. Israel, Phys. Rev. {\bf D41} (1990) 1796.}
\lref\ori{A. Ori,  Phys. Rev. Lett. {\bf 67} (1991) 789; {\bf 68} (1992) 2117.}
\lref\ghdm{G. Horowitz and D. Marolf, Phys. Rev. {\bf D55} (1997) 835,
hep-th/9605224.}
\lref\ghd{G. Horowitz and D. Marolf, Phys. Rev. {\bf D55} (1997) 846,
hep-th/9606113.}
\lref\bek{J. Bekenstein, gr-qc/9605059.}
\lref\lawi{F. Larsen and F. Wilczek, Phys. Lett. {\bf B375} (1996) 37, 
hep-th/9511064; Nucl. Phys. {\bf B475} (1996) 627, hep-th/9604134;
hep-th/9609084.}
\lref\cvts{M. Cvetic and A. Tseytlin, Phys. Rev. {\bf D53} (1996) 5619,
hep-th/9512031; A. Tseytlin, Mod. Phys. Lett. {\bf A11} (1996) 689,
hep-th/9601177;  Nucl. Phys. {\bf B477} (1996) 431, hep-th/9605091.}

\lref\kmr{N. Kaloper, R. Myers and H. Roussel, hep-th/9612248.}
\lref\ascv{A. Strominger and C. Vafa, Phys. Lett. {\bf B379} (1996) 99,
hep-th/9601029.}
\lref\homa{G. Horowitz and D. Marolf, hep-th/9610171.}
\lref\pol{J. Polchinski, Phys. Rev. Lett. {\bf 75} (1995) 4724,
hep-th/9510017.}
\lref\mtw{C. Misner, K. Thorne, and J. Wheeler, {\it Gravitation}, Sec. 32.6
(W. H. Freeman, New York, 1973).}
\lref\gar{D. Garfinkle and T. Vachaspati, Phys. Rev. {\bf D42} (1990) 1960;
D. Garfinkle, Phys. Rev. {\bf D46} (1992) 4286.}
\lref\host{G. Horowitz and A. Steif, Phys. Rev. Lett. {\bf 64} (1990) 260.}
%-------------------
% title page
%-------------------
%
\baselineskip 16pt
\Title{\vbox{\baselineskip12pt
\line{\hfil   UCSBTH-97-01}
\line{\hfil \tt hep-th/9701077} }}
{\vbox{
{\centerline{Black Strings and Classical Hair}}
}}
\centerline{\ticp Gary T. Horowitz\footnote{}{\ttsmall
gary@cosmic.physics.ucsb.edu, yangh@cosmic1.physics.ucsb.edu}
and Haisong Yang}
\bigskip
\vskip.1in
\centerline{\it Department of Physics, University of California,
Santa Barbara, CA 93106, USA}
\bigskip
\centerline{\bf Abstract}
\bigskip
We examine the geometry near the event horizon of a
family of black string solutions
with traveling waves. It has previously been shown that the metric is 
continuous there. Contrary to expectations, we find that the geometry is not
smooth, and the horizon 
becomes singular whenever a wave is present. Both five dimensional and
six dimensional black strings are considered with similar results.

\Date{January, 1997}

\newsec{Introduction}

More than twenty five years ago Wheeler proposed that ``black holes have no hair"
which captured the idea 
that in gravitational collapse, most of the information about
what forms the black hole is not available classically in the external 
black hole solution. 
This was supported by the uniqueness theorem
for the Kerr-Newman solution in Einstein-Maxwell theory, and various extensions
of this result showing that the solution remains unique when simple additional
matter fields are included.
It was latter realized that more complicated matter fields can lead
to new black hole solutions \more. But the  spirit of the ``no hair conjecture"
was preserved  in that the new solutions contained only a few additional
parameters which represented only a small amount of information about the
initial state which
formed the black hole. For a recent review see \bek.

Recently, a potentially more serious violation of the ``no hair conjecture"
has been proposed \lawi. In string theory, one naturally considers a charged
black hole which is extended in some internal direction, so the solution
describes a black string. It was shown
that in the extremal limit, one can add waves to this solution  traveling
along the string. Unlike most gravitational waves, these waves do not
radiate to infinity or fall into the horizon: They appear to be a form
of classical hair.  Since the  waves are not characterized by a few 
parameters, but instead by arbitrary functions, it was suggested that 
perhaps all the information about the black string could be contained in
these waves on the exterior geometry \refs{\lawi,\cvts}.

In \ghdm, the geometry near the event horizon of a black string with
traveling waves was examined. It was shown that the metric was at least
$C^0$ there, which was sufficient to insure that the horizon area was
well defined. It was expected at the time that the horizon was smooth, 
and it was just the complicated nature of the solution that made it difficult to
find good coordinates everywhere. Here we
examine this geometry  more closely and show that the curvature 
actually {\it diverges} at the horizon
whenever the wave is present\foot{We assume that the direction along the
black string
is compact, as required by an ``internal" direction. If one relaxes this 
assumption, some waves are nonsingular.}.  We will see that
this singularity is null
in the sense that
all scalar curvature invariants remain finite, and rather mild since
the total tidal distortion remains finite. Nevertheless, since the classical
``no hair conjecture" usually assumes a regular event horizon, these waves
cannot be viewed as examples of classical hair. 

There are two different examples of black strings with traveling waves
which have been discussed. One is a six dimensional black string which reduces
to a five dimensional black hole, and the other is a five dimensional
black string reducing to a four dimensional black hole. There are also 
different types of waves which can be added to these black strings. 
In section 2, we examine two different waves on 
the six dimensional black string and in section 3 we repeat the analysis
for the five dimensional black string. In both cases we show that the
horizon becomes singular whenever the wave is present. In section 4 we show
that the total tidal distortion experienced by an observer falling into the
singularity remains finite, which leaves open the possibility that one can
pass through this region. Section 5 contains a brief discussion.

\newsec{Curvature of Six Dimensional Black String
 With Traveling Waves }

We start with the low energy effective action of Type IIB 
string theory in the ten dimensional Einstein frame,
 keeping only the metric, the dilaton and RR
three form H:

\eqn\action{
 	S={1 \over {16 \pi G}}
	\int d^{10} x \sqrt{- g} \( {R- {1 \over 2}(\nabla \phi )^2
 	- {1 \over 12} e^{\phi} H^2}\)
}
A class of extremal black string solutions to
\action\ with  traveling
waves were studied in  \refs{\ghdm, \ghd}. The metric
for these solutions takes the form:

\eqn\sxwvmetric{\eqalign{
ds^2_{10} =& - \( 1-{ r_0^2 \over r^2} \)dudv +  \[
{p(u) \over r^2} - 2\(1 - { r_0^2 \over r^2}\)\ddot{f}_a(u) y^a
  \]du^2  \cr 
 &+ \(1 -{ r_0^2 \over r^2}\)^{-2} dr^2
+ r^2 d \Omega_3^2 + dy_ady^a}}
where $y^a\ (a=1,2,3,4)$ are  periodic coordinates on an internal $ T^4$.
 $u = t-z$ and $ v = t+z$ where $z$ is a coordinate on an $S^1$ and is identified
with period $L$, so $p(u)$ and $f_a(u)$ are periodic 
functions with period $L$. The dot represents  a derivative with respect
to $u$. When both $p(u)$ and $f_a(u)$ are constant, the metric is stationary and
represents the product of $T^4$ and a six dimensional extremal black string.
There is an event horizon at $r= r_0$ which can be shown to be
smooth. The functions $p(u)$ and $f_a(u)$ describe waves traveling along
this black string, which we will call longitudinal and internal waves
respectively.

We begin by considering the case when only the
longitudinal wave is present. In this case,
the problem is reduced to six  dimensions. 
Let us first compute the curvature in the  natural coordinate
system, i.e., the coordinates
$(u,v,r, \theta, \phi, \psi)$. We will set $r_0=1$
for the rest of this section.
Keeping only the 
longitudinal wave, we can write the 
the metric as 
\eqn\sxpwvmetric{
ds_6^2 = - \( 1-{ 1  \over r^2} \)dudv +{ p(u) \over
r^2}du^2 + \(1 - { 1 \over r^2}\)^{-2} dr^2
+ r^2 d \Omega_3^2 }

We choose the following tetrad to calculate 
the curvature components:
\eqn\sxnatetrad{\eqalign{
\omega_1 =& rd\theta \cr
\omega_2 =& r\sin\theta d\varphi \cr
\omega_3 =& r\sin\theta \sin\varphi d\psi \cr
\omega_4 =& \(1 - { 1 \over r^2}\)^{-1} dr  \cr
\omega_5 =& { p^{1/2}(u) \over r}du - {r^2-1 \over {2rp^{1/2}(u)}}dv \cr
\omega_6 =& {r^2-1 \over {2rp^{1/2}(u)}}dv 
}}
so $ ds^2 = \omega_1^2 + \omega_2^2 + \omega_3^2 + \omega_4^2
+ \omega_5^2 - \omega_6^2$. In this tetrad, 
 the curvature components turn out to be
\eqn\sxnacurvature{\eqalign{
&R_{1212} = R_{2323} = R_{3131} = {2 r^2 -1 \over r^6}  \cr 
&R_{1414} = R_{2424} = R_{3434} =   {-2r^2+2 \over r^6} \cr
&R_{1515} = R_{2525} = R_{3535} =  {(r^2-1)^2 \over r^6} \cr 
&R_{1616} = R_{2626} = R_{3636} = {r^4-1 \over r^6} \cr
&R_{1516} = R_{2526} = R_{3536} = {r^2-1 \over r^4} \cr
&R_{4545} = {-3r^4+6r^2-4 \over r^6} \cr
&R_{4546} = {-3r^2+3 \over r^4} \cr
&R_{4646} = { -3r^4+4 \over r^6} \cr
&R_{5656} = {1 \over r^6}
}}
The rest of the components are zero except the ones that can be obtained
by symmetry. Notice that all curvature components are finite at the
horizon $r=1$. In fact, they are all independent of $p(u)$! This is a 
special property
of the tetrad \sxnatetrad\ that we have chosen. But it follows that
all scalar curvature invariants 
are independent of $p(u)$ and finite at the horizon.
The Ricci tensor takes a very simple form:
\eqn\sxRicci{\eqalign{
&R_{11} = R_{22} = R_{33} = {2 \over r^6} \cr
&R_{44} = -{2 \over r^6}  \cr
&R_{55} = -{2 \over r^6}  \cr
&R_{66} = {2 \over r^6}
}} 

At first sight, it appears that these results show that the curvature 
is completely
well behaved as one approaches the horizon. 
But one must realize that the (future) event horizon
 lies not only at $r=1$, but also at $u,v=+\infty$ 
and the coordinate system $(u,v,r)$ becomes ill-defined 
 at the horizon. More importantly, the
tetrad \sxnatetrad\ also becomes ill-defined at the horizon and is
related to a good tetrad by an infinite boost involving 
$ \omega_4,\omega_5, \omega_6$. 
 The black string curvature components \sxnacurvature\ are not
invariant under a general boost, so potential divergences may arise. 
(It is interesting to notice that the $4,5,6$ components of the
Ricci tensor \sxRicci\ are  proportional to the three dimensional Minkowski
metric and hence are boost invariant. This is consistent with the fact
that the Ricci tensor does not change when a traveling wave is added \gar.)
 For  the $p(u)= constant$ case (no traveling
wave), an infinite boost is still necessary, but turns out not to cause a
problem.
 An analytic extension through  the horizon
 has been found in the Appendix
of \ghdm\ and the  horizon is smooth. For the 
general case, $p(u)$ oscillates an infinite number of times before
reaching the horizon and this new feature causes  
  the curvature singularity at the 
horizon, as we will see.   

A coordinate system in which the black string metric \sxwvmetric\
 is $C^0$ at the horizon was found in 
\ghdm, although only the leading order terms near the horizon were given
explicitly.
The exact metric (for the case of a longitudinal wave) can be written as
\eqn\sxcnodmetric{\eqalign{
ds^2 =& -W^2dUdV + \sigma^2W^4 \[ (r^2-1)(4r^4-3) \]
dU^2 \cr
&+\[ 2\sigma W^4r^2 (r^2-1) (2r^2+1)+ 6W^2\(\int_0^U 
\sigma^2 W^4 dU\) \]dqdU \cr
&+W^4r^6dq^2 + r^2d\Omega_3^2
}}
where the coordinate transformations are given by
\eqn\sxcotranf{\eqalign{
&\sigma^2(u)+\dot{\sigma}(u) = p(u) \cr
&G(u) = e^{\int_0^u\sigma du} \cr
&U =- \int_u^{+\infty}{du \over G^2}  \cr
&W = G {\( 1-{1 \over r^2}\)}^{1/2} \cr
&q =-{1 \over 2W^2} - 3 \int_0^U\sigma dU \cr
&V = v -{ \sigma(u) \over r^2-1} - 2 \int_0^u\sigma^2 du
+ 3 \int_0^U \sigma^2 W^2 dU 
}}
Some motivation for these transformations can be found in \ghdm\ (where
$V$ was called $\nu$).
$\sigma(u)$ is a periodic function with the same 
 period $ L$ as $p(u)$. $U$ is a function of $u$ and is one of the
 new coordinates.
The (future) event horizon lies at $U=0$ and $U<0$ 
corresponds to the region outside the horizon.
(The above
definition of $U$ only holds outside the horizon. Inside
the horizon, $U$ is defined to be  $U = \int_{-\infty}^u{du \over G^2}$.)
$r$ and $W$ are functions of
$U$ and $q$ and can be obtained explicitly
by inverting \sxcotranf. It follows that $r=1$ when $U=0$.
The metric is independent
of $V$, so $\partial/\partial V$ is a null Killing vector.
The last term in the definition of $V$ is an integral over $U$ at
constant $q$. It contributes $3 \sigma^2 W^2 dU + 6 (\int_0^U \sigma^2 W^4 dU)
dq$ to $dV$.

The 
horizon area is $A=2\pi^2\int_0^L\sigma du$ \ghdm. 
Notice that we can actually choose different wave profiles
$p_{in}(u),p_{out}(u)$ for the region inside and outside 
the horizon as long as they satisfy a matching condition:
\eqn\matchcd{
\int_0^L\sigma_{in}(u)du = \int_0^L\sigma_{out}(u)du
}
i.e., the horizon area has to match.

The metric \sxcnodmetric\ is only $C^0$ at $U=0$ since $\sigma$ is a periodic
function of $u$ and $U=0$ corresponds to $u= \infty$. So $\sigma$ oscillates
an infinite number of times near the horizon and does not have a well defined
limit. 
For simplicity, we will choose the following null tetrad
to calculate the curvature components
\eqn\sxcnodtetrad{\eqalign{
\omega_1 =&rd\theta \cr
\omega_2 =& r\sin\theta d\varphi \cr
\omega_3 =& r\sin\theta \sin\varphi d\psi \cr
\omega_4 =& W^2r^3dq \cr
\omega_5 =& W^2dU \cr
\omega_6 =& -{1 \over  2}dV + {1 \over  2}\sigma^2W^2
	\[(r^2-1)(4r^4-3)\]dU \cr
  &           +\[\sigma W^2 r^2(r^2-1)(2r^2+1)+ 3 \(\int_0^U
\sigma^2 W^4 dU\)\]dq 
}}
so $ds^2= \omega_1^2+ \omega_2^2 +\omega_3^2+ \omega_4^2
+2\omega_5\omega_6.$ This is a good tetrad for the entire region,
including the horizon.  
The curvature components turn out to be
\eqn\sxcnodcurvature{\eqalign{
&R_{1212} = R_{2323} = R_{3131} = {2r^2-1 \over r^6} \cr
&R_{1414} = R_{2424} = R_{3434} =  {-2r^2+2\over r^6} \cr
&R_{1415} = R_{2425} = R_{3435} = {-2r^2+3 \over r^3 }\sigma \cr
&R_{1515} = R_{2525} = R_{3535} = {\dot\sigma \over r^2-1} 
				+ O(\dot\sigma, \sigma) \cr
&R_{1516} = R_{2526} = R_{3536} = {-r^2+1 \over r^6} \cr
&R_{4545} = - {3\dot\sigma \over r^2-1}+O(\dot\sigma, \sigma) \cr
&R_{4546} =  {3r^2-4 \over r^6} \cr
&R_{4556} = {6r^2-9 \over r^3}\sigma \cr
&R_{5656} = { 1 \over r^6}
}}
The rest of the components are zero except the ones that 
can be obtained by symmetry. The above expressions are valid
everywhere, with $O(\dot\sigma, \sigma)$ denoting terms which are
finite at the horizon ($r=1$).
The components $ R_{1515}$ , $R_{2525}$ , $ R_{3535}$ , $R_{4545}$
clearly diverge at the horizon and the geometry is singular.
An observer crossing the horizon 
would feel infinite tidal force.
The curvature diverges when $\lim_{u\rightarrow \infty} \dot \sigma(u) 
\ne 0$. This will hold whenever $p(u)$ is not constant, since it is periodic
and $\sigma$ is defined as in \sxcotranf. 
This
calculation of the curvature is reliable even though the metric is only $C^0$
at the horizon, 
for the following reason. The metric is analytic everywhere away from the
horizon  and in this region we can calculate the curvature components
in either tetrad \sxnatetrad\ or tetrad \sxcnodtetrad. As one approaches
the horizon, one physically wants to compute the curvature in a frame which
is parallelly propagated along a geodesic. When checking for curvature
singularities, it suffices to use a tetrad which is continuous
across the horizon.
The $C^0$ metric is sufficient for determining continuous tetrads.
Calculating the curvature components in tetrad \sxcnodtetrad\
is equivalent to boosting the curvature components calculated in
\sxnatetrad\ and an infinite boost 
is needed as we approach the horizon. 

We have also calculated the curvature when both longitudinal and internal
waves are present. The result is very similar. In the continuous tetrad, 
the divergent components are of the form ${\dot\sigma(u) 
\over r^2-1}$ and ${\ddot f_a(u) \over r^2-1}$. As before,
all scalars remain finite at the horizon. 

There are two other types of traveling waves that were discussed  
 in \ghdm\ and \ghd. They  represent the oscillation of the black
 string  in the four
 external spatial directions and the angular momentum
 of the black string. We did not calculate the curvature for
these two types of waves but we expect the result to be similar.

\newsec{Curvature Of  Five Dimensional Black String 
With Traveling Waves}
 
Type IIA string theory admits a class of solutions describing  
five dimensional extremal black string with 
traveling waves. The metric takes the 
 form
\eqn\fvwvmetric{\eqalign{
ds_{10}^2 =& - \( 1-{r_0 \over r}\)dudv+ \[ {p(u) 
\over r} - 2\(1-{r_0 \over r}\) \ddot{f}_a(u)y^a 
  \]du^2 \cr
& + \(1-{r_0 \over r} \)^{-2}dr^2 +r^2d\Omega^2+dy_ady^a
}}
The coordinates are similar to the six dimensional case except that
there are now five periodic $y^a$ coordinates parameterizing an internal
$T^5$.
As before, we refer to the   
 waves described by $p(u)$ and $ \ddot{f}_a(u)$ 
as longitudinal and internal waves. 

All the results of the five  dimensional black string will  be
very similar to the previous section.
We first consider the case where only the
longitudinal wave exists and comment on other types 
 of waves later.

When there is only a longitudinal wave, 
the problem is reduced to five  dimensions.
Let us first compute the curvature in the natural
coordinate system  $(u,v,r, \theta,\phi )$.
Keeping only the longitudinal wave and setting $r_0 =1$,
we can write the metric as 
\eqn\fvpwvmetric{\eqalign{
ds_5^2 = - \(1-{1 \over r}\)dudv+ {p(u) \over r}du^2
+\(1-{1 \over r}\)^{-2} dr^2  +r^2d\Omega^2 
}}
We choose the following tetrad to calculate the curvature components:
\eqn\fvnatetrad{\eqalign{
\omega_1 =& rd\theta \cr
\omega_2 =& r\sin\theta d\varphi \cr
\omega_3 =& \(1-{1 \over r}\)^{-1}dr  \cr
\omega_4 =& \[ {p(u) \over r}\]^{1/2}du - {r-1 \over 
 2[p(u)r]^{1/2}}dv \cr
\omega_5 =& {r-1 \over 2[ p(u)r]^{1/2}}dv
}}
so $ds^2 = \omega_1^2 +\omega_2^2 +\omega_3^2 + \omega_4^2
-\omega_5^2$. In this tetrad, the nontrivial curvature components are
\eqn\fvnacurvature{\eqalign{
&R_{1212}= {2r-1 \over r^4} \cr
&R_{1313}= R_{2323}= {-r+1 \over r^4} \cr
&R_{1414}= R_{2424} = {(r-1)^2 \over 2r^4} \cr
&R_{1415} = R_{2425} = {r-1 \over 2r^3} \cr
&R_{1515} = R_{2525} = {r^2-1 \over 2r^4} \cr
&R_{3434} = {-4r^2+8r-5 \over  4r^4} \cr
&R_{3435} = {-r+1 \over r^3} \cr
&R_{3535} = {-4r^2+5 \over 4r^4} \cr
&R_{4545} = {1 \over 4 r^4} 
}}

Once again, all curvature components are independent of $p(u)$
 and 
all  scalars that can be formed out of curvature
tensor and the metric are finite at horizon.
The Ricci tensor takes a simple form
\eqn\fvRicci{\eqalign{     
R_{11}=& {1 \over r^4} \cr
R_{22} =& {1 \over r^4} \cr
R_{33} =&- { 1 \over 2r^4} \cr
R_{44} =&- {1 \over 2r^4} \cr
R_{55} =& { 1 \over 2r^4} 
}}

All curvature components seems to be  well behaved at the 
horizon. 
However, similar to the six dimensional case, the (future) event 
horizon lies at $r=1,\ u,v =+\infty $ and the tetrad \fvnatetrad\
becomes ill-defined at the horizon. It is again related to a 
good tetrad by an infinite boost. For the $p(u) = constant$
case, an analytic extension through the horizon can be found. 
For the general case, $p(u)$ oscillates an infinite number of times before 
reaching the horizon and this causes a curvature singularity there.

By copying the steps in the 
six dimensional black string case, we can find new coordinates in which
the metric is $C^0$ at the horizon. The metric then takes the form 
 
\eqn\fvcnodmetric{\eqalign{
ds^2 =&   -{2W^2 \over r}dUdV + 4\sigma^2W^4
 { (r-1)(9r^2+3r-2)  \over r } dU^2  \cr
&  +\( 8\sigma W^4 {(r-1)(3r^2+2r+1) \over r}
+ {48W^2 \over r}  \int_0^U\sigma^2 W^4dU\)dUdq
+4 W^4r^2dq^2  + r^2d\Omega^2
}}
where the coordinate transformations are given by 
\eqn\fvcotranf{\eqalign{
&\sigma^2(u)+2\dot\sigma(u)= p(u) \cr
&G(u)= e^{{1 \over 2}\int_0^u\sigma(u)du} \cr
&U = -{1 \over 2} \int^{+\infty}_u{du \over G^2} \cr
&W=G(r-1)^{1/2} \cr
&q=-{1 \over 2W^2}-3\int_0^U\sigma dU \cr
&V = v  - {2\sigma \over r-1} - 3 \int_0^u\sigma^2du 
+12\int_0^U\sigma^2W^2 dU 
}}
$\sigma(u)$ is a periodic function with period $L$. The
event horizon lies at $U=0$ and $U<0$ corresponds to the 
region outside the horizon. As in 
the  six dimensional case, we can choose different wave profiles
$p_{in}(u), p_{out}(u)$ for regions inside and outside
the horizon as long as the horizon area is matched. The metric
is analytic both inside and outside the horizon but is 
only continuous at the horizon. We will calculate 
the curvature components in the  null tetrad
 
\eqn\fvcnodtetrad{\eqalign{
\omega_1 =& rd\theta \cr
\omega_2 =& r\sin\theta d\varphi \cr
\omega_3 =& 2W^2r dq \cr
\omega_4 =& {2 W^2 \over r}dU \cr
\omega_5 =& -{1\over 2}dV+ \sigma^2W^2(r-1)(9r^2+3r-2)dU \cr
	    &+\[2\sigma W^2(r-1)(3r^2+2r+1) +12\( \int_0^U\sigma^2
W^4dU\)\]dq
}}
so $ds^2 = \omega_1^2 +\omega_2^2+\omega_3^2+2\omega_4\omega_5$.
The  nontrivial components of the curvature are

\eqn\fvcnodcurvature{\eqalign{
&R_{1212}= {2r-1 \over r^4} \cr
&R_{1313}= R_{2323} = {-r+1 \over r^4} \cr
&R_{1314} = R_{2324} = {-3r+4 \over 2r^2}\sigma \cr
&R_{1414}= R_{2424} = { \dot\sigma \over r-1} + O(\sigma, \dot\sigma)  \cr
&R_{1415} = R_{2425} = {-r+1 \over 2r^4} \cr
&R_{3434} =- {2\dot\sigma \over r-1}+O(\sigma, \dot\sigma)  \cr
&R_{3435} = {4r-5 \over 4r^4} \cr
&R_{3445} = {3r-4 \over r^2}\sigma \cr
&R_{4545} = {1 \over 4r^4} 
}}

The components $R_{1414}, R_{2424}$ and $ R_{3434}$ clearly diverge at 
the horizon $r=1$. 
We also calculated the curvature in the case where both longitudinal and 
internal waves are present. The divergent components are of the
form ${\dot\sigma(u) \over r-1}$ and ${\ddot{f}_a(u) \over r-1}$ and all scalars
remain finite at the horizon.

There is another type of traveling wave that was discussed 
in \ghd. It represents the oscillation of the black string
in the three external spatial directions.
    We did not calculate the curvature for this  case 
but we expect the result to be similar.

\newsec{ The Curvature Components Are  Integrably Finite
At The Event Horizon}

 A free falling object feels a tidal force  which is proportional to
the curvature components
$R_{i\tau j\tau}$, where $\tau$,
$i\ (i=1,2,...,D-1)$ denote components in an orthonormal frame carried 
by the observer.  
For the black string solutions with traveling waves discussed in this paper,
these components diverge 
at the horizon. However in this section we show that 
the curvature components have 
the property that they are integrably finite in 
the sense that for a free-falling observer, these components integrated twice
over proper time are finite
at the horizon.  
This  in turn means that the total distortion on an object is finite,
and a small object  can
survive the tidal force when crossing the horizon. We will explain 
what  the criterion for ``small" is shortly.

The fact that the integrated tidal force remains finite is closely related
to the fact that the metric is well defined and continuous at the horizon.
It is also  reminiscent of the
singularity at the Cauchy horizon inside a charged or rotating  black
hole \refs{\pois, \ori}.

 Let us first show
that the curvature components are integrably finite.
Consider the six dimensional black string case. Near the horizon, we have

\eqn\curestone{|R_{i\tau j\tau}|={C_1 \over r^2-1}
=C_2{G^2 \over W^2} = C_3 G^2 }
where $C_1,C_2,C_3$ are finite constants.
The second and third steps can be seen 
from equation \sxcotranf. 
Next, let us define
\eqn\trade{\eqalign{
&\sigma_0\equiv {1 \over L}\int_0^L\sigma du, \cr
&G_0(u) \equiv e^{\int_0^u\sigma_0 du} = e^{\sigma_0 u}, \quad
U_0 \equiv -\int^{+\infty}_u {du \over G_0^2}=-{1 \over 2\sigma_0 G_0^2}, \cr
&\eta_G \equiv {G \over G_0}, \qquad
\eta_U \equiv {U \over U_0}. 
}} 
 $\eta_G$ is a periodic function with period $L$
and therefore has a finite value bounded from below
and above. From this, one can show that  $\eta_U$  
  is also bounded from below and above. So, $G^2
= {C \over U}$ and
\eqn\curexttwo{|R_{i\tau j\tau}|={C_4 \over U}}
Finally, let us
notice that $ U$ is a  null coordinate and ${dU
\over d\tau}$ is approximately a finite constant near
the horizon. This leads to
\eqn\curestthree{|R_{i\tau j\tau}|={C_5 \over \tau}}
$C, C_4, C_5$ are all finite constants and 
$\tau$ is the proper time measured from 
the horizon. 

Thus, the curvature components integrated twice over
proper time are clearly finite at the horizon. 
This is also true for the five  dimensional black 
string.   

As an object falls into the black string, the 
tidal force gets bigger and bigger. We can divide 
this process into two periods. The first is when 
the stress caused by the tidal force is smaller 
than the   
maximal value that the object can resist. After 
the transition point, the stress caused by the tidal force  
 exceeds  this maximal value
and we can think of the object as  made of dust-like point particles
that follow their own geodesics. The criterion 
of surviving the tidal force
should be that the distortion has to be much 
smaller than the size of the object when it
 reaches the horizon. 

 At any time, there is a stress 
distribution within the object.  
The maximal stress  
$T_{max}$ can be estimated as \mtw:
\eqn\stress{
T_{max} \sim \rho l^2 |R_{i\tau j\tau}|
}
where $\rho$ is the mass  density of the object and
$l$ is the linear size of the object. For a small
object, the transition point will be close 
to the horizon. Given this and the fact that the curvature
components are integrably finite, it is not 
difficult to see that the distortion will be
small compared with the size of the object and 
therefore it will reach  the horizon intact.

\newsec{Discussion}

We have seen that the addition of traveling waves to extremal black strings
causes the horizon to become singular. This supports the ``no hair conjecture"
and reinforces the idea that solutions with regular horizons (having 
compact cross-sections) are characterized
by only a few parameters. Nevertheless, the waves should not be discarded
as totally unphysical. There are indications that these waves may still play
a role 
in string theory. First, because of the null translational symmetry and
special properties of the spatial geometry, these solutions are likely to
be exact, and not receive $\alpha'$ 
corrections even though the curvature becomes large \refs{\cvts,\host}.

Second, in \refs{\ghdm,\ghd}
it was shown that the waves traveling along an extremal black string 
affect the area of the event horizon as well as the distribution of momentum
along the string. Using recent results in string theory \refs{\pol, \ascv}, 
it was also shown that one could reproduce the 
Bekenstein-Hawking entropy of this black string by counting states in
weakly coupled string theory with the same charge and momentum distribution.
This is true despite the fact that the curvature at the horizon diverges.

It should perhaps be noted that traveling waves can be added {\it inside} the
horizon of a black string without causing the horizon to become singular
\homa. In fact there are a large class of such modes, which may explain
where information is stored inside the horizon.

We have considered the case where the direction along the black string
is compact. If one considers the case where it is noncompact, then it
is clear from the results in sections 2 and 3 that nonsingular waves are
possible. One can simply choose $p(u)$ to be constant for large $u$ or
approach a constant sufficiently quickly. Then 
$\dot \sigma$ vanishes for large $u$ fast enough so that
all curvature components remain finite. However
it is not clear whether these waves should be called classical hair.
Since $u = t-z$, an asymptotic observer at fixed $z$ cannot detect the
wave at late times. In this sense, the ``hair" is transitory. This is
not a problem when $z$ is periodic.

\vskip 1cm
After this work was completed, \kmr\ appeared which has some overlap with the
results presented here.

 \vskip 1cm
 \centerline{\bf Acknowledgments}
 It is a pleasure to thank D. Marolf for discussions.
 This work was supported in part by NSF Grant PHY95-07065.

\listrefs
\end